\documentclass[11pt,a4paper,oneside]{article}
\usepackage[T1]{fontenc}
\usepackage[ansinew]{inputenc}
\usepackage[english]{babel}
\usepackage{amsfonts}
\usepackage{bm}
\usepackage{booktabs}
\usepackage{array}
\usepackage{amsthm}
\usepackage{amssymb}
\usepackage{graphicx}
\usepackage{braket}
\usepackage{verbatim}
\usepackage[dvipsnames]{xcolor}
\usepackage{caption}
\usepackage{textcomp}
\usepackage{url}
\usepackage{float}
\usepackage{leftindex}

\usepackage{multicol}

\usepackage{setspace}
\usepackage{graphicx} 
\usepackage{graphics} 
\usepackage{float} 
\usepackage{subcaption}
\usepackage{amsmath}
\usepackage{amssymb}
\usepackage{amsthm}
\usepackage{mathtools}
\usepackage{enumerate}
\usepackage{enumitem}
\usepackage[colorlinks=true, allcolors=black]{hyperref}
\usepackage{tensor}
\usepackage{tikz}

\DeclareMathOperator\erf{erf}
\DeclareMathOperator\F{F}

\numberwithin{equation}{section}
\usepackage[symbol]{footmisc}

\begin{document}
\title{\bf Coherent electrically-charged quantum black holes} 
\author{
Tommaso~Antonelli$^{1}$\thanks{Email: t.antonelli@sussex.ac.uk} ,
Marco~Sebastianutti$^{1}$\thanks{Email: m.sebastianutti@sussex.ac.uk} ,
Andrea~Giusti$^{2,3}$\thanks{Email: andrea.giusti9@unibo.it} 
\\
\\
$^1${\em Department of Physics and Astronomy, University of Sussex}\\ 
{\em Brighton, BN1 9QH, United Kingdom}
\\
\\
$^2${\em DIFA \& ${\cal AM}^2$, University of Bologna, 40126 Bologna, Italy}
\\
\\
$^3${\em I.N.F.N., Sezione di Bologna, I.S.~FLAG}
\\
{\em viale B.~Pichat~6/2, 40127 Bologna, Italy}
}
\maketitle

\begin{abstract}
\noindent We improve upon the results presented in [R.~Casadio, {\em et al.}, Phys. Rev. D \textbf{105} (2022) 124026] deriving a quantum-corrected Reissner--Nordstr\"{o}m geometry containing an integrable singularity at its center while being devoid of spurious oscillations around the classical configuration. We further investigate some relevant physical observables, related to geodesics and quasinormal modes of scalar perturbations, associated with this geometry to complement our theoretical analysis.
\end{abstract}

\newpage
\renewcommand{\thefootnote}{\arabic{footnote}}

\section{Introduction}
The Reissner--Nordstr\"{o}m (RN) spacetime is an exact solution of the Einstein--Maxwell equations describing a static spherically symmetric electrically charged black hole. Notably, in General Relativity a uniqueness theorem holds~\cite{Israel:1967za} stating that the RN black hole is the only static and asymptotically flat {\em electrovacuum} black hole solution of Einstein's equations. Realistic astrophysical black holes are expected to carry a negligible electric charge at the end of the gravitational collapse, yet RN black holes are still of significant mathematical interest due to their rich causal structure (see~\cite{Hawking:1973uf}). Specifically, a RN black hole contains a curvature singularity at $r=0$, similarly to many other classical black hole solutions, and an inner (Cauchy) horizon. Note that the Cauchy horizon is known to imply an instability under perturbations~\cite{PhysRevLett.63.1663,PhysRevD.41.1796} via the mass-inflation phenomenon.

The resolution of the problem represented by the central curvature singularity has been one of the underlying motivations for the quest for a quantum theory of gravity. This led to the development of a large number of regular black hole spacetimes (see e.g.~\cite{Bambi:2023try}, and references therein) which are meant to describe effective geometries that could potentially emerge from quantum gravity. Interestingly, it turns out that regular black holes, although being devoid of curvature singularities, usually contain inner Cauchy horizons which could still lead to instabilities. 

The {\em coherent state approach to quantum black holes}~\cite{Casadio:2016zpl,Casadio:2017cdv,Casadio:2021eio} aims at addressing the issue of singularities in a way that significantly differs from the standard lore of regular black holes. The formalism associated with this approach relies on some simple quantum-mechanical considerations and it is not meant to capture all the properties of a full quantum theory of gravity. Instead, this approach is meant to provide a simple and formal implementation of the {\em quantum $N$-portrait of black holes} formulated by G.~Dvali and C.~Gomez in~\cite{Dvali:2011aa}. Within this scheme, black holes are treated as macroscopic high-multiplicity quantum states of both gravitons and matter fields, and the coherent state approach then represents a sort of ``mean-field approach'' to such systems.

This work is organized as follows: in Sec.~\ref{sec-RN} we improve upon the results of~\cite{Casadio:2022ndh} obtaining a quantum-corrected RN metric, through the coherent state approach, devoid of spurious oscillations around the classical geometry. In Sec.~\ref{sec:source} we compute the effective stress-energy tensor for our quantum-corrected geometry. In Sec.~\ref{sec-hor} we investigate the causal structure of the new geometry determining the location of inner and event horizons as a function of the parameters of the model. In Sec.~\ref{sec-sing} we analyze the nature of the singularity at $r=0$. Notably, the singularity can be removed for a specific choice of the coherent-state UV regulator or, otherwise, it is still present but of much milder nature (integrable singularity) compared to the classical one. In Sec.~\ref{sec-geo} we study observables related to the geodesics on this spacetime. In Sec.~\ref{sec-qnm} we compute the scalar quasinormal modes of the proposed quantum-corrected RN geometry. Finally, in Sec.~\ref{sec-conclusions} we provide some concluding remarks.

\section{Improving the quantum-corrected Reissner--Nordstr{\"o}m spacetime}
\label{sec-RN}
In this section we discuss how to smooth-out the spurious oscillations of the quantum RN geometry derived in~\cite{Casadio:2022ndh} by means of the regularization procedure presented in~\cite{Feng:2024nvv}. Since the discussion presented here will follow closely those in~\cite{Casadio:2022ndh} and~\cite{Feng:2024nvv}, 
we will not reproduce the full computation of the quantum-corrected RN ``potential''. Instead, we will focus on detailing the key differences introduced in this work.

According to the framework of {\em coherent quantum black holes}~\cite{Casadio:2016zpl,Casadio:2017cdv,Casadio:2021eio}, the emerging classical geometry of a black hole is obtained as the expectation value of a quantum metric tensor operator over a suitable quantum state. Specifically, if we consider a generic static spherically-symmetric asymptotically flat classical black hole geometry
\begin{equation}
\label{eq:fmetric}
    ds^2=-f(r)dt^2+\frac{dr^2}{f(r)}+r^2d\Omega^2, \quad f(r) := 1+2V(r) \, ,
\end{equation}
(where $d\Omega^2$ denotes the line element on $S^2$ and $r$ being the areal radius) with the locations of the horizons belonging to the zeroes of $f(r)$, within this formalism, we have that the classical potential $V$ is conveniently described as the expectation value of a free massless scalar field $\widehat{\Phi}$ over a coherent state $\ket{g}$, i.e.
\begin{equation}
\label{condition}
    \braket{g \, | \, \widehat{\Phi} \, | \, g } = V \, .
\end{equation}
Said scalar field is meant to capture the non-perturbative collective behavior of the relevant degrees of freedom required to reproduce the classical geometry in the full quantum theory, irrespective of its microscopic description. For further details, we refer the reader to~\cite{Casadio:2021eio,Feng:2024nvv}. 

In this work we are interested in the emergence of the classical RN geometry, i.e.
\begin{equation}\label{eq:RNpot}
    V_{\rm RN}(r)=-\frac{G_{\rm N} M}{r}+\frac{G_{\rm N} Q^2}{2r^2} \, ,
\end{equation}
where $M$ and $Q$ are respectively the Arnowitt-Deser-Misner (ADM) mass and the electric charge of the black hole.

As proven in~\cite{Casadio:2022ndh}, there cannot exist a coherent state of a free massless scalar field that is both normalizable and satisfies Eq.~\eqref{condition} for the RN potential~\eqref{eq:RNpot}. This is due to the fact that the contribution of modes of arbitrary large momentum induce a ultraviolet (UV) divergence in the occupation number of the would-be coherent state $\ket{g}$. This problem can be resolved by introducing a sharp UV cut-off $R_{\rm s}$ identified with the size of the quantum core of the system. In other words, the quantum-corrected potential is obtained from the regularized expectation value~\cite{Casadio:2022ndh}
\begin{equation}
\label{eq:oldRN}
\begin{split}
\leftindex^{\rm (old)\;}V^{\rm q}_{\rm RN}(r) :=&  \braket{g \, | \, \widehat{\Phi} \, | \, g }_{\rm reg} =
\int_{0}^{R_{\rm s}^{-1}} \frac{k^2 dk}{2\pi^2} \, \tilde{V}_{\rm RN}(k) \, j_0(kr)\\
=& -\frac{G_{\rm N}M}{r}\frac{2}{\pi}\,
{\rm Si}\!\left(\frac{r}{R_{\rm s}}\right)
+
\frac{G_{\rm N}\,Q^2}{2\,r^2}
\left[
1
-\cos\!\left(\frac{r}{R_{\rm s}}\right)
\right]\, ,
\end{split}
\end{equation}
where $\tilde{V}_{\rm RN}(k)$ denotes the Fourier transform of the classical RN potential, that reads
\begin{equation}
    \tilde{V}_{\rm RN}(k)=-\frac{4\pi G_{\rm N}M}{k^2}+\frac{\pi^2 G_{\rm N}Q^2}{k} \, ,
\end{equation}
while ${\rm Si} (z)$ denotes the sine-integral function.

The oscillations induced by the sine-integral and cosine functions in~\eqref{eq:oldRN} are spurious and carry no physical significance. Specifically, they are due to the sharp nature of the cut-off $R_{\rm s}$ appearing in the regularized potential. 

\begin{figure}
	\centering   \includegraphics[width=0.6\textwidth]{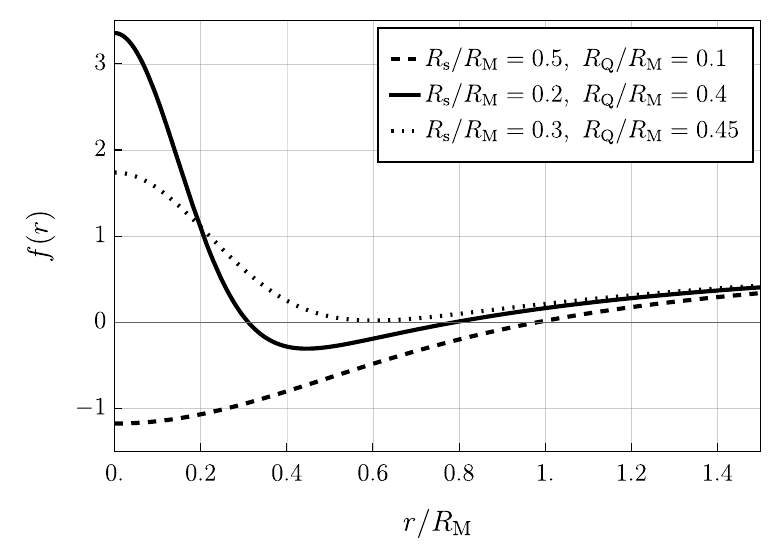}
    \captionsetup{width=0.9\textwidth, font=footnotesize}
	\caption{Metric function $f(r)=1+2V^{\rm q}_{\rm RN}(r)$ for different values of the parameters $R_{\rm s}$ and $R_{\rm Q}$ (in units of $R_{\rm M}$).}
	\label{fig:f}
\end{figure}

As argued in~\cite{Feng:2024nvv} for the quantum-corrected Schwarzschild geometry, these spurious oscillations can be removed by regularizing the expectation value in~\eqref{eq:oldRN} with a Gaussian regulator related to $R_{\rm s}$, rather than using this 
scale as a sharp UV cut-off. In other words the {\em improved quantum-corrected RN potential} reads 
\begin{equation}
    \begin{split}
    V_{\rm RN}^{\rm q}(r) :=& \int_0^\infty \frac{k^2 dk} {2\pi^2}\left[\tilde{V}_{\rm RN}(k) \, {\rm e}^{-\frac{k^2R_{\rm s}^2}{4}}\right]j_0(kr) \\
    =& -\frac{G_{\rm N} M}{r}\erf{\hspace{-1mm}\left(\frac{r}{R_{\rm s}}\right)}+\frac{G_{\rm N} Q^2}{2r^2}\frac{2r}{R_{\rm s}}\F\hspace{-1mm}\left(\frac{r}{R_{\rm s}}\right) \, .
    \end{split}
\end{equation}
with $\erf(z)$ and $\F(z)$ being respectively the error and Dawson functions:
\begin{equation}
  \erf(x)=\frac{2}{\sqrt{\pi}}\int_0^x e^{-t^2}dt,\qquad \F(x)=e^{-x^2}\int_0^x e^{t^2} dt.
\end{equation}
The quantum-corrected metric function hence reads
\begin{equation}
    \label{eq:quantumf}
    f(r)= 1+2V_{\rm RN}^{\rm q}(r)=1-\frac{R_{\rm M}}{r}\erf{\hspace{-1mm}\left(\frac{r}{R_{\rm s}}\right)}+\frac{R_{\rm Q}^2}{r^2}\frac{2r}{R_{\rm s}}\F\hspace{-1mm}\left(\frac{r}{R_{\rm s}}\right),
\end{equation}
where $R_{\rm M}\equiv2G_{\rm N}M$ and $R_{\rm Q}\equiv\sqrt{G_{\rm N}}Q$, see Fig.~\ref{fig:f}.

Clearly, deviations from the classical RN spacetime emerge in the region where $r$ becomes comparable to $R_{\rm s}$, while they are quickly suppressed at larger radii (as can be easily checked by means of the known asymptotic expansions of $\erf(z)$ and $\F(z)$). Furthermore, it is immediate to see that if the core radius $R_{\rm s}\to0^+$, i.e. if we restore the singularity at the center of the RN black hole, we recover the classical RN geometry.

\section{Effective source}\label{sec:source}
As for the classical RN case, the metric~\eqref{eq:fmetric} with the quantum-corrected metric function~\eqref{eq:quantumf} does not correspond to any vacuum solutions of the Einstein equations. Additionally, the effective stress-energy tensor sourcing the quantum-corrected RN metric is significantly more involved due to the ``smoothing'' of the potential near $r=0$. Specifically, simple computations yield
\begin{equation}
    \tensor{T}{^\mu_\nu}=\frac{\tensor{G}{^\mu_\nu}}{8\pi G_{\rm N}}=\text{diag}\left(-\rho^{\rm q},p_r^{\rm q},p_t^{\rm q},p_t^{\rm q}\right),
\end{equation}
where the energy density $\rho^{\rm q}$, radial pressure $p^{\rm q}_r$, and tangential pressure $p^{\rm q}_t$ for the quantum-corrected metric function~\eqref{eq:quantumf} are given by
\begin{align}\label{eq:rho}
    \rho^{\rm q}(r)=-p^{\rm q}_{r}(r)&=\frac{1-f(r)-rf'(r)}{8\pi G_{\rm N}\,r^2}\notag\\
    &=\rho_{\rm M}\frac{R_{\rm s}^2}{r^2}e^{-\frac{r^2}{R_{\rm s}^2}}-2\rho_{\rm Q}(r)\frac{r^2}{R_{\rm s}^2}\left(1-\frac{2r}{R_{\rm s}}\F\hspace{-1mm}\left(\frac{r}{R_{\rm s}}\right)\right),
\end{align}
and 
\begin{align}\label{eq:pt}
    p^{\rm q}_t(r)&=\frac{2f'(r)+rf''(r)}{16\pi G_{\rm N}\,r}\notag\\
    &=\rho_{\rm M}e^{-\frac{r^2}{R_{\rm s}^2}}-2\rho_{\rm Q}(r)\frac{r^5}{R_{\rm s}^5}\left[\frac{R_{\rm s}}{r}-\left(2-\frac{R_{\rm s}^2}{r^2}\right)\F\hspace{-1mm}\left(\frac{r}{R_{\rm s}}\right)\right],
\end{align}
with 
$$\rho_{\rm M}\equiv\frac{M}{2\pi^{3/2} R_{\rm s}^3} \quad \mbox{and} \quad\rho_{\rm Q}(r)\equiv\frac{Q^2}{8\pi r^4} \, .$$ 

\begin{figure}[ht!]
    \centering 
    \includegraphics[width=0.48\textwidth]{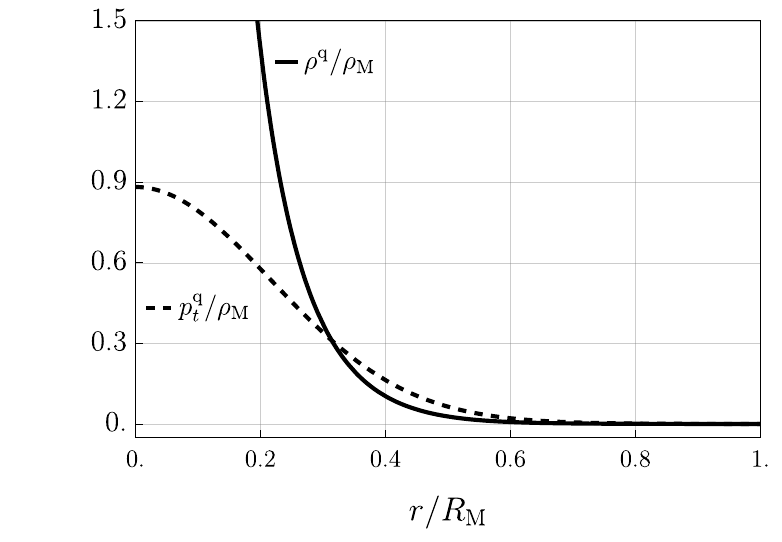}
    \hspace{0.5cm}\includegraphics[width=0.48\textwidth]{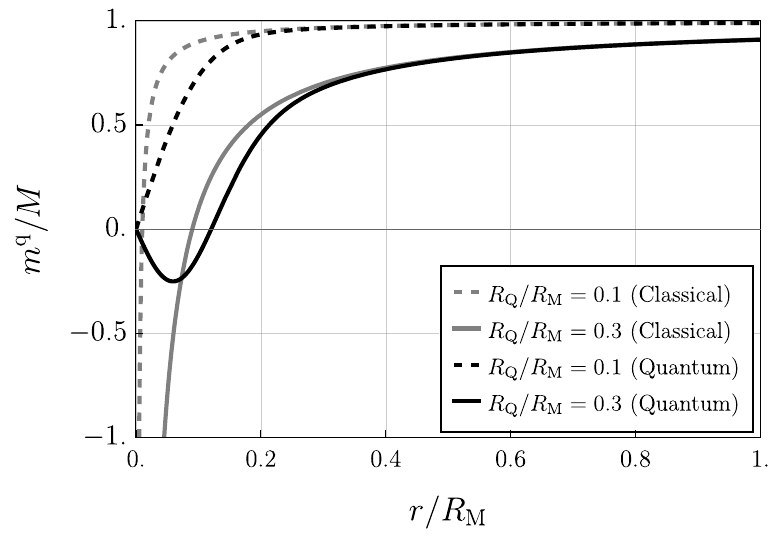}
    \captionsetup{width=0.9\textwidth, font=footnotesize}
    \caption{Left: normalized effective energy density (solid line) and tangential pressure (dashed line) for $R_{\rm s}/R_{\rm M}=0.3$ and $R_{\rm Q}/R_{\rm M}=0.1$. Right: normalized mass function for $R_{\rm s}/R_{\rm M}=0.1$.}
    \label{fig:SETM}
\end{figure}

It is now worth observing that, for $r/R_{\rm s}\gg1$ the above expressions asymptotically approach $\rho_{\rm Q}(r)$, while in the limit for $r\to0^+$ one has that
\begin{equation}
    \rho^{\rm q}(r)=-p^{\rm q}_{r}(r)\simeq\left(\rho_{\rm M}-2\rho_{\rm Q}(R_{\rm s})\right)\frac{R_{\rm s}^2}{r^2}+\mathcal{O}(1),\qquad p_{t}^{\rm q}(r)\simeq\rho_{\rm M}-4\rho_{\rm Q}(R_{\rm s})+\mathcal{O}(r^2) \, ,
\end{equation}
i.e. the energy density and radial pressure {\em diverge} at $r=0$ (unless $\rho_{\rm M}=2\rho_{\rm Q}(R_{\rm s})$) while the tangential pressure remains finite, see Fig.~\ref{fig:SETM}. Nonetheless, the volume integrals of $\rho^{\rm q}$, $p^{\rm q}_r$, and $p^{\rm q}_t$ are always finite since these functions are {\em locally integrable} for $r>0$ with respect to the 3-volume measure. This point will be important for the classification of the singularity at $r=0$.

Notably, from Eq.~\eqref{eq:rho} we can compute the Misner--Sharp mass~\cite{Misner:1964je,Faraoni:2015ula} for our quantum-corrected geometry, that yields
\begin{equation}\label{eq:mr}
    m^{\rm q}(r):= 4\pi \int_0 ^r \rho^{\rm q}(x) \, x^2 \, dx =-\frac{r\,V_{\rm RN}^{\rm q}(r)}{G_{\rm N}}=M \erf{\hspace{-1mm}\left(\frac{r}{R_{\rm s}}\right)}-\frac{Q^2}{R_{\rm s}}\F\hspace{-1mm}\left(\frac{r}{R_{\rm s}}\right),
\end{equation}
which, for $r\to+\infty$, yields the ADM mass of the system $M$; while it reduces to the classical RN quasilocal mass as $R_{\rm s}\to 0^+$, i.e.
\begin{equation}
    \lim_{R_{\rm s}\to 0^+}m^{\rm q}(r)=M-\frac{Q^2}{2r} \, .
\end{equation}

\section{Inner and outer horizons}
\label{sec-hor}
The classical RN spacetime features two horizons whose location is determined by
\begin{equation}
	f(R_{\pm})=1+2V_{\rm RN}(R_{\pm})=0,
\end{equation}
where
\begin{equation}
	R_{\pm}=\frac{R_{\rm M}}{2}\pm\frac{1}{2}\sqrt{R_{\rm M}^2-4R_{\rm Q}^2}.
\end{equation}
In the non-extremal case, i.e.~for $R_{\rm M}>2R_{\rm Q}$, $R_{+}$ and $R_{-}$ are respectively the event and Cauchy horizons. 

\begin{figure}[ht!]
	\centering
	\captionsetup{width=0.9\textwidth, font=footnotesize}
	\includegraphics[width=0.6\textwidth]{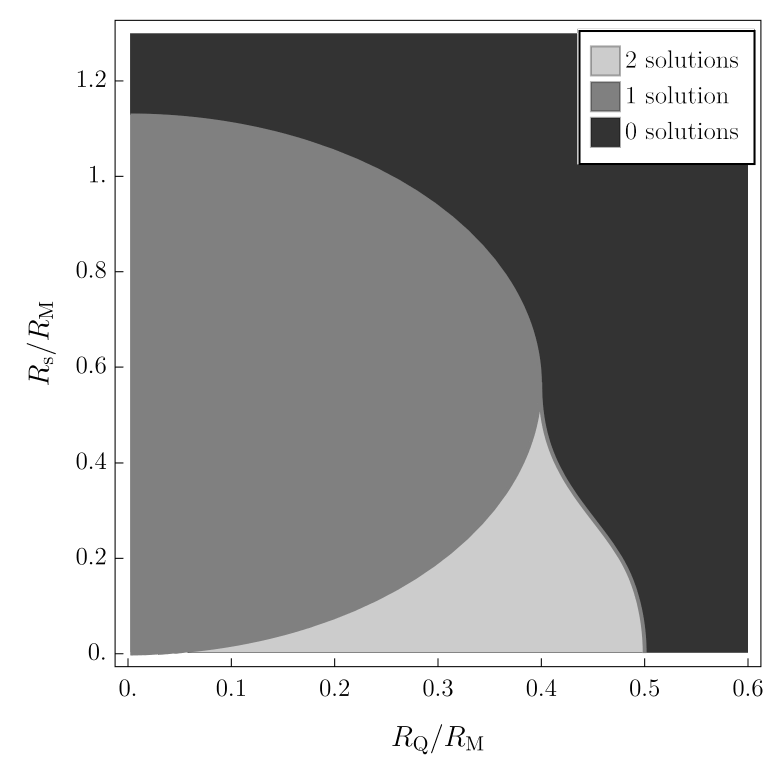}
	\caption{Number of solutions to the equation $V_{\rm RN}^{\rm q}(r)=-1/2$, or equivalently $f(r)=0$, in the parameter space $R_{\rm s}$--$R_{\rm Q}$ (in units of $R_{\rm M}$). Inside the elliptical dark gray region, the equation $f(r)=0$ has one solution with multiplicity one; while on the dark gray boundary between the light gray region (two solutions) and the black region (no solutions), $f(r)=0$ admits one solution with multiplicity two, corresponding to the extremal case.
    }
	\label{fig:horizons_RN}
\end{figure}

The quantum-corrected RN spacetime can also present both outer and inner horizons whose locations are given by the solutions of
\begin{equation}\label{eq:horizons}
	V_{\rm RN}^{\rm q}(R_{\pm}^{\rm q})=-\frac{1}{2} \, .
\end{equation}
Clearly, $R_{\pm}^{\rm q}$ deviate from $R_{\pm}$ due to the contribution of the quantum core of radius $R_{\rm s}$ (where we use the convention $R_{-}^{\rm q}<R_{+}^{\rm q}$ in the non-extremal case). Unlike their classical counterparts, the expressions for $R_{\pm}^{\rm q}$ cannot be attained analytically due to the transcendental nature of Eq.~\eqref{eq:horizons}; however, bounds on some of the black hole parameters $R_{\rm M}$, $R_{\rm Q}$, and $R_{\rm s}$ can be determined and employed to guide the numerical approximation of the solutions of Eq.~\eqref{eq:horizons}. In general, we have three possible scenarios: (i) two distinct horizons (non-extremal case, $R_{-}^{\rm q}<R_{+}^{\rm q}$); (ii) one horizon (either corresponding to a multiplicity one solution of Eq.~\eqref{eq:horizons}; or with multiplicity two, namely the extremal case, $R_{-}^{\rm q}=R_{+}^{\rm q}$); (iii) horizon-less geometry (Eq.~\eqref{eq:horizons} has no solution). The various cases are summarized in Fig.~\ref{fig:horizons_RN}. 

From this discussion we can expect that the Carter-Penrose diagrams would not differ from those of Schwarzschild [case (ii) with multiplicity one] or Reissner--Nordtr\"om [case (i), (ii) with multiplicity two, and (iii)], if we trust the classical notion of geodesics to extend past $R_{\rm s}$. Taking the effective classical geometry described here at face value then implies that inner horizons, when present, are true Cauchy horizons and the geometry can suffer of the standard mass inflation instability \cite{Poisson:1989zz}. However, in order to know the behavior of infalling matter and light inside the quantum core we would need a precise description of matter and gravity in the full quantum gravity regime inside the core.

In order to determine the number of solutions to $V_{\rm RN}^{\rm q}(r)=-1/2$, given that
\begin{equation}
  \lim_{r\to\infty}V_{\rm RN}^{\rm q}(r)=0 \, ,  
\end{equation}
it is useful to determine under which conditions we obtain $V_{\rm RN}^{\rm q}(0)=-1/2$. From this latter equation we obtain a quadratic in $R_{\rm s}/R_{\rm M}$, which has roots
\begin{equation}
	\lambda_{1}^\pm=\frac{1}{\sqrt{\pi}}\left(1\pm\sqrt{1-2\pi\frac{R_{\rm Q}^2}{R_{\rm M}^2}}\right).
\end{equation}
If we fix a certain $R_{\rm Q}/R_{\rm M}$ such that $R_{\rm Q}/R_{\rm M}<1/\sqrt{2\pi}\simeq0.399$, we have that $\lambda^\pm_1$ are real and we can have all three situations depending on the value of $R_{\rm s}/R_{\rm M}$. By varying this parameter, the values $\lambda_{1}^\pm$ give us the threshold values at which the geometry varies from having two horizons to one, and from one horizon to none. In this case the quantum-corrected geometry has: two horizons if $R_{\rm s}/R_{\rm M}<\lambda_{1}^-$; one horizon if $\lambda_{1}^-<R_{\rm s}/R_{\rm M}<\lambda_{1}^+$; no horizons if $R_{\rm s}/R_{\rm M}>\lambda_{1}^+$.

Given these results we can observe that, from a geometric viewpoint, the dark gray region in Fig.~\ref{fig:horizons_RN} is described by an ellipse of cartesian equation
\begin{equation}
	2\pi\frac{R_{\rm Q}^2}{R_{\rm M}^2}+\left(\pi\frac{R_{\rm s}}{R_{\rm M}}-1\right)^2=1.
\end{equation}

If $1/\sqrt{2\pi}<R_{\rm Q}/R_{\rm M}<1/2$ we can only have either two horizons or none, 
represented by the light gray and the black region in Fig.~\ref{fig:horizons_RN} respectively. Note that, while crossing the boundary between the light gray and black regions we have a one-dimensional subset of the parameter space with coincident (doubly degenerate) solutions of $V_{\rm RN}^{\rm q}(r)=-1/2$, i.e. the case with a single horizon (extremal scenario). This boundary region is highlighted in Fig.~\ref{fig:horizons_RN} with a dark gray line. To further characterize the boundary between the light gray and black regions we would have to solve the equations $f(r)=f'(r)=0$ (so that the horizon is doubly degenerate) to obtain an expression of $R_{\rm s}/R_{\rm M}$ as a function of $R_{\rm Q}/R_{\rm M}$ (or vice versa). This cannot be done analytically, in general, 
however we can still compute a parametric form of such a solution. To this end, we can solve $f(r)=f'(r)=0$ for $r=R^{\rm q}_+=R^{\rm q}_-=t\, R_{\rm s}$, where $t\in(0,\infty)$ is a parameter that relates the size of the single (doubly degenerate) horizon to the radius of the core. This allows us to solve the equations $f(r)=f'(r)=0$ for $R_{\rm s}/R_{\rm M}$ and $R_{\rm Q}/R_{\rm M}$ leaving the dependence on $t$ explicit. 
Hence, the boundary between the light gray and black regions in Fig.~\ref{fig:horizons_RN} is given by the plane curve $\gamma(t)$ that reads
\begin{equation}
\begin{split}
  \left(\frac{R_{\rm s}}{R_{\rm M}},\frac{R_{\rm Q}}{R_{\rm M}}\right)&\equiv\gamma(t)=(\gamma_1(t),\gamma_2(t))=\\
  &=\Biggl(\dfrac{2\F(t)+e^{t^2}\sqrt{\pi}(-1+2t\F(t))\erf(t)}{e^{t^2}\sqrt{\pi}\left[-t+(2t^2+1)\F(t)\right]},\\
  &\hspace{0.82cm} \dfrac{\sqrt{\left[-2t+e^{t^2}\sqrt{\pi}\erf(t)\right]\left[2\F(t)+e^{t^2}\sqrt{\pi}(-1+2t\F(t))\erf(t)\right]}}{e^{t^2}\sqrt{2\pi}\left[-t+(2t^2+1)\F(t)\right]}\Biggr).
\end{split}
\end{equation}

A physical requirement that we may impose on the solution is that the quantum core remains hidden inside the event (outer) horizon, i.e. $R_{\rm s}<R_{+}^{\rm q}$. To understand whether the core radius sits outside or inside the event horizon, we first have to determine in which cases $R_{\rm s}$ can coincide with one of the horizons. This means determining the solutions of $V_{\rm RN}^{\rm q}(R_{\rm s})=-1/2$, which is a quadratic in $R_{\rm s}/R_{\rm M}$ and gives us two roots
\begin{equation}
    \lambda_{2}^\pm=\frac{\erf(1)}{2}\left(1\pm\sqrt{1-\frac{8\F(1)}{\erf^2(1)}\frac{R_{\rm Q}^2}{R_{\rm M}^2}}\right).
\end{equation}
If $R_{\rm Q}/R_{\rm M}\le\gamma_2(1)\simeq0.404$, then it is necessary and sufficient to impose $R_{\rm s}/R_{\rm M}<\lambda^+_2$ to have that $R_{\rm s}<R_{+}^{\rm q}$, while if $R_{\rm Q}/R_{\rm M}>\gamma_2(1)$, no additional bounds on $R_{\rm s}/R_{\rm M}$ are needed and $R_{\rm s}$ is always smaller than $R_{+}^{\rm q}$, if the latter exists. 

From Fig.~\ref{fig:horizons_RN} one can also point out that
$\gamma_2(t)<{1}/{2}$, $\forall t\in(0,\infty)$, which immediately tells us that for the extremal case the Weak Gravity Conjecture bound\footnote{The asymptotic corrections of $V^{\rm q}_{\rm RN}(r)$ to the classical potential $V_{\rm RN}(r)$ are ${\cal O} (r^{-4})$ as $r\to\infty$, so the parameters $M$ and $Q$ coincide with the total Komar mass and charge of the black hole.}~\cite{NimaArkani-Hamed_2007,Rudelius:2024mhq}:
\begin{equation}
  G_{\rm N}^{-1/2}\,\frac{Q}{M}=2\frac{R_{\rm Q}}{R_{\rm M}}=1-\epsilon, \quad \epsilon>0 \, ,
\end{equation}
is satisfied, thus suggesting that charged coherent quantum black holes, as those discussed here, could belong to the class of solutions of some viable UV-complete gravity theories~\cite{Antonelli:2025zwc}.

\section{Singularity or regularity at the origin}
\label{sec-sing}
Since the quantum-corrected RN metric function $f(r)=1+2V_{\rm RN}^{\rm q}(r)$ in Eq.~\eqref{eq:quantumf} is analytic, it can be expanded at $r=0$ as
\begin{equation}
	f(r)=\sum_{k=0}^\infty f_k\, r^k,
\end{equation}
where $f_k=f^{(k)}(0)/k!$. Then, the curvature invariants as $r \to 0^+$ read
\begin{equation}
  R=-\frac{2(f_0-1)}{r^2}-\frac{6f_1}{r}-12f_2+\mathcal{O}(r),\hspace{6.82cm}
\end{equation}
\begin{equation}
\begin{split}
	\tensor{R}{^\mu_\nu}\tensor{R}{^\nu_\mu}=\ &\frac{2(f_0-1)^2}{r^4}+\frac{8f_1(f_0-1)}{r^3}+\frac{2\left(6f_2(f_0-1)+5f_1^2\right)}{r^2}+\frac{4\left(4f_3(f_0-1)+9f_1f_2\right)}{r}\\
	&+4\left(5f_4(f_0-1)+14f_1f_3+9f_2^2\right)+\mathcal{O}(r),
\end{split}
\end{equation}
\begin{equation}
\begin{split}
    \mathcal{K}=\ &\frac{4(f_0-1)^2}{r^4}+\frac{8f_1(f_0-1)}{r^3}+\frac{8\left(f_2(f_0-1)+f_1^2\right)}{r^2}+\frac{8\left(f_3(f_0-1)+3f_1f_2\right)}{r}\hspace{3.5mm}  \\
	&+8\left(f_4(f_0-1)+4f_1f_3+3f_2^2\right)+\mathcal{O}(r),
\end{split}
\end{equation}
where $R$ denotes the Ricci scalar, $\tensor{R}{^\mu_\nu}\tensor{R}{^\nu_\mu}$ is the ``square'' of the Ricci tensor, and $\mathcal{K} = \tensor{R}{^{\mu\nu\rho \sigma}}\tensor{R}{_{\mu\nu\rho \sigma}}$ denotes the Kretschmann scalar.

Clearly, from the above expressions, we have that the considered invariants are all finite for $f_0=1$ and $f_1=0$. Specializing to the quantum-corrected RN metric function, with $f(r)$ as in Eq.~\eqref{eq:quantumf}, we have that
\begin{equation}
	f_0=1+\frac{2}{R_{\rm s}}\left(\frac{R_{\rm Q}^2}{R_{\rm s}}-\frac{R_{\rm M}}{\sqrt{\pi}}\right), \qquad f_1=0, \qquad f_2=-\frac{4}{3R_{\rm s}^3}\left(\frac{R_{\rm Q}^2}{R_{\rm s}}-\frac{R_{\rm M}}{2\sqrt{\pi}}\right) \, .
\end{equation}

We can now distinguish two scenarios:
\begin{itemize}
    \item If $f_0=1$, which corresponds to
\begin{equation}
  \label{eq:regular_condition}
	R_{\rm s}=R_{\rm s}^{*}\equiv\frac{\sqrt{\pi}R_{\rm Q}^2}{R_{\rm M}} \, ,
\end{equation}
depending on the ratio $R_{\rm Q}/R_{\rm M}$, it is possible to obtain a configuration that is regular at $r=0$ and features a single (doubly degenerate) horizon. Looking at Fig.~\ref{fig:horizons_RN}, this spacetime is realized at the intersection of the parabola identified in Eq.~\eqref{eq:regular_condition} and the boundary $\gamma(t)$ between the light gray and black regions.
\item If $f_0\neq1$, i.e. $R_{\rm s}\neq R_{\rm s}^{*}$, the quantum-corrected RN geometry contains an {\em integrable singularity}~\cite{Lukash:2013ts} at $r=0$. Specifically, although curvature invariants diverge, their divergence is milder than that of the RN geometry and the volume integral of the effective densities and pressure is finite (as already pointed out in Sec.~\ref{sec:source}).\footnote{Note that for $R_{\rm Q}=0$, the curvature divergences at $r=0$ cannot be removed unless we also set $R_{\rm M}=0$, which corresponds to Minkowski spacetime.} Note that, although these kind of singularities might potentially allow for low-regularity~\cite{Clarke:1994cw} extensions of the spacetime past them, they are not free of potential physical shortcomings, as recently pointed out in~\cite{Arrechea:2025fkk}.
\end{itemize}
Note that these results mimic what was originally determined in~\cite{Casadio:2022ndh} for the quantum-corrected RN spacetime obtained with the sharp UV cut-off. In other words, the UV regularization does not seem to affect the nature of the singularity at $r=0$.

\section{Geodesics}
\label{sec-geo}
We can now apply the standard approach for investigating geodesics on static spherically symmetric spacetimes (see e.g.~\cite{Wald:1984rg}).

Consider the motion of a point-like particle in the geometry determined by~\eqref{eq:fmetric}, working on the equatorial plane ($\theta = \pi / 2$) without loss of generality. Then, the geodesic equation reads  
\begin{equation}
    \dot r^2=E^2-V_{\rm eff}(r) \, , \qquad V_{\rm eff}(r)=f(r)\left(
  \frac{L^2}{r^2}-\mu\right) \, ,
\end{equation}
with $\mu$ determining the norm of the tangent vector to a geodesic (taking values $-1$, $0$ or $+1$ depending on whether the geodesic is timelike, null, or spacelike respectively), the dot denoting the derivative with respect to the affine parameter along the geodesic, $E = f(r) \, \dot t = {\rm const.}$ the ``energy'' of the particle along the geodesic, and $L = r^2 \,\dot\phi = {\rm const.}$ the ``angular momentum'' of the particle along the geodesic.

\begin{figure}
\centering
\captionsetup{width=0.9\textwidth, font=footnotesize}
    \includegraphics[width=0.49 \textwidth]{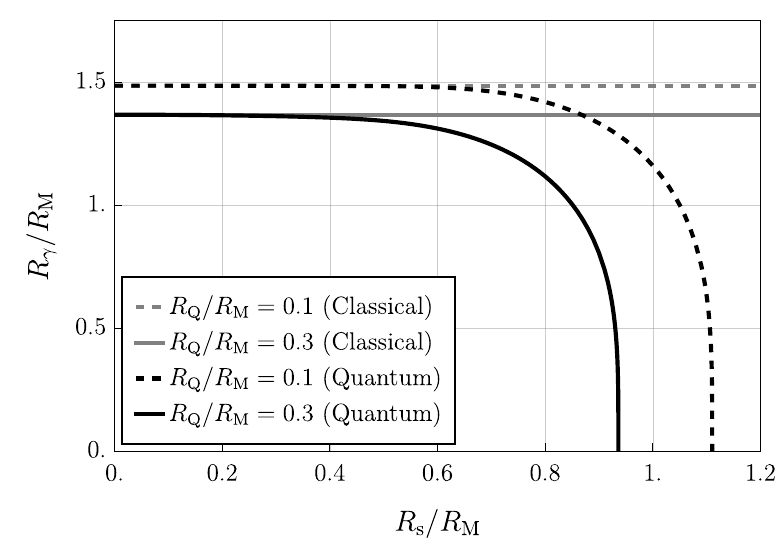}
\includegraphics[width=0.49\textwidth]{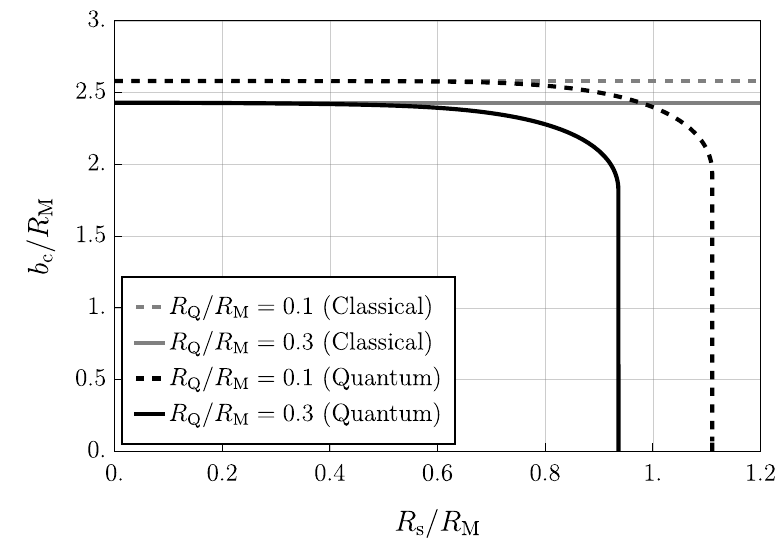}
\caption{Left: (outer) photon radius $R_\gamma$ (in units of $R_{\rm M}$) as a function of $R_{\rm s}/R_{\rm M}$. Right: critical impact parameter $b_{\rm c}$ (in units of $R_{\rm M}$) as a function of $R_{\rm s}/R_{\rm M}$. For both plots we have chosen to depict two values of $R_{\rm Q}/R_{\rm M}$, and we always compare their quantum-corrected values to the corresponding classical RN cases ($R_{\rm s}\to0^+$).
}
\label{fig:photon_rad_crit_impact}
\end{figure}

As pointed out in~\cite{Urmanov:2024qai} for the quantum-corrected Schwarzschild geometry, several quantities of interest can be calculated from this equation. For example we can calculate the photon ring $R_{\gamma}$, which is defined as the radius of circular orbits for massless particles ($\mu=0$), i.e. $V_{\rm eff}'(R_{\gamma})=0$. Hence, $R_{\gamma}$ satisfies the condition 
\begin{equation}
\label{eq:RGammaCondition}
  R_{\gamma}\, f'(R_{\gamma})-2\,f(R_{\gamma})=0
\end{equation}
For the classical RN black hole ($R_{\rm s}=0$), Eq.~\eqref{eq:RGammaCondition} has two solutions given by
\begin{equation}
  \label{eq:photon_rad_RN}
  R^{\rm RN}_\gamma=\frac{3R_{\rm M}}{4}\left(1\pm\sqrt{1-\frac{32 R_{\rm Q}^2}{9R_{\rm M}^2}}\right)
\end{equation}
In our general case, where $R_{\rm s}\neq 0$, we would need to solve the following equation for $R_\gamma$:
\begin{equation}
  1-\frac{R_{\rm Q}^2}{R_{\rm s}^2}+\frac{R_{\rm M}}{\sqrt{\pi}R_{\rm s}}\exp\hspace{-1mm}\left({-\frac{R_{\gamma}^2}{R_{\rm s}^2}}\right)+\frac{R_{\rm Q}^2(3R_{\rm s}^2+2R_{\gamma}^2)}{R_\gamma R_{\rm s}^3}\F\hspace{-1mm}\left(\frac{R_{\gamma}}{R_{\rm s}}\right)-\frac{3R_{\rm M}}{2R_\gamma}\erf\hspace{-1mm}\left(\frac{R_\gamma}{R_{\rm s}}\right)=0
\end{equation}
which can only be done via numerical methods, see Fig.~\ref{fig:photon_rad_crit_impact}, left panel.

Once the photon radius $R_\gamma$ is determined, one can also compute the so-called critical impact parameter $b_{\rm c}$, which is defined as
\begin{equation}
\label{eq:critical_param}
b_{\rm c}
:=
\frac{R_\gamma}{\sqrt{f(R_\gamma)}}
\ ,
\end{equation}
and this quantity represents the minimal ratio $b \equiv L/E$ (the impact parameter) below which the particle falls into the black hole.

For a classical RN solution ($R_{\rm s}=0$), if we take the plus sign in Eq.~\eqref{eq:photon_rad_RN}, we get that $b_{\rm c}$ can be expressed as 
\begin{equation}
  b_{\rm c}=\frac{3\sqrt{3}}{4\sqrt{2}}\,\frac{\left(1+\sqrt{1-\dfrac{32R_{\rm Q}^2}{9R_{\rm M}^2}}\right)^2}{\,\sqrt{1-\dfrac{8R_{\rm Q}^2}{3R_{\rm M}^2}+\sqrt{1-\dfrac{32R_{\rm Q}^2}{9R_{\rm M}^2}}}}\,R_{\rm M} \, .
\end{equation}
For the geometry determined by the metric function~\eqref{eq:quantumf} with $R_{\rm s} \neq 0$ we no longer have an analytical expression for $b_c$, and again we have to resort to numerical computations, see again Fig.~\ref{fig:photon_rad_crit_impact}, right panel.

\begin{figure}
\centering
\captionsetup{width=0.9\textwidth, font=footnotesize}
\includegraphics[width=0.49\textwidth]{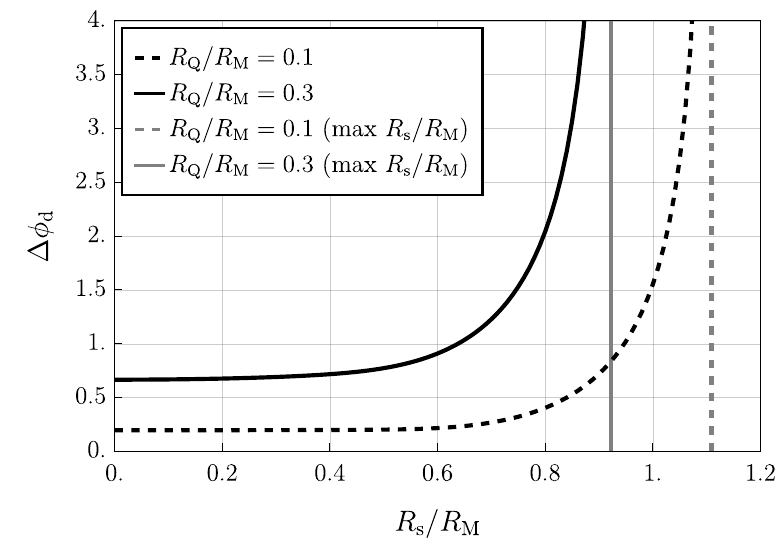}
\includegraphics[width=0.49\textwidth]{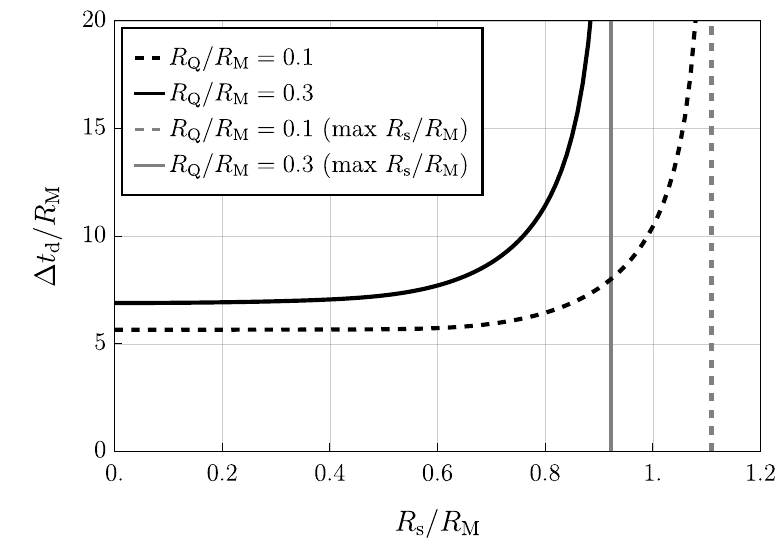}
\caption{Left: $\Delta\phi_{\rm d}$ as a function of $R_{\rm s}/R_{\rm M}$. Right: Shapiro time delay $\Delta t_{\rm d}/R_{\rm M}$ as a function of $R_{\rm s}/R_{\rm M}$ where, for simplicity, we have taken $r_{A}=r_{B}=2r_0$, $r_0$ being the radius associated to $b$. For both plots we have chosen two values of $R_{\rm Q}/R_{\rm M}$ with $b=2R_{\rm M}$. Moreover, we have depicted in gray the vertical asymptotes corresponding to the maximum values of $R_{\rm s}$, for which $b=b_{\rm c}$. It is also worth noting that the values of the observables for the RN case are represented by the horizontal asymptotes approached by the curves as $R_{\rm s} \to 0^+$.}
\label{fig:phi_time}
\end{figure}

Another important observable quantity is the gravitational lensing of light rays. Specifically, considering the case of a light ray ($\mu=0$) we can derive the following relation from the geodesic equations:
\begin{equation}
  \label{eq:phi_rad_geod}
  \dot\phi=\pm\frac{\dot r}{r\sqrt{\dfrac{r^2}{b^2}-f(r)}}
\end{equation}
where, again, $b\equiv L/E$. The trajectory has a turning point at radius $r=r_0$, and at this point we get $\dot r=0$. Since at $r=r_0$ we have that $\dot\phi\neq 0$, from Eq.~\eqref{eq:phi_rad_geod} we can conclude that 
\begin{equation}
\label{eq:brel}
  b=\frac{r_0}{\sqrt{f(r_0)}} \, ,
\end{equation}
thus showing the relation between the impact parameter $b$ and the minimal radius $r_0$ in the trajectory. We can now compute the change in angle $\Delta \phi$ due to the presence of the black hole, and to this end we integrate Eq.~\eqref{eq:phi_rad_geod} assuming that the light ray comes from spatial infinity, reaches $r_0$, and drifts off towards spatial infinity; this yields
\begin{equation}
  \Delta \phi=2\int_{r_0}^\infty \frac{dr}{r\sqrt{\dfrac{r^2}{b^2}-f(r)}}\,.
\end{equation}
The deflection angle $\Delta \phi_{\rm d} = \Delta \phi-\pi$ due to the presence of the black hole cannot be evaluated analytically. For numerical results concerning $\Delta \phi_{\rm d}$ for our quantum-corrected RN geometry see Fig.~\ref{fig:phi_time}, left panel.

Lastly, another interesting observable is provided by the Shapiro time delay (see e.g.~\cite{Gasperini:2013cru}), i.e.~the time difference of a light pulse due to the presence of a gravitational object near the trajectory of the signal. 
Following a similar procedure to the one that led to Eq.~\eqref{eq:phi_rad_geod}, we can derive the following relation from the geodesic equations:
\begin{equation}
  \label{eq:time_radius_geod}
  \dot t=\pm\frac{\dot r}{f(r)\sqrt{1-f(r)\,\dfrac{b^2}{r^2}}}\,.
\end{equation}
Consider now an object $A$ located at $r=r_A$ emitting a light signal directed to a second object $B$ at $r=r_B$; and assume that $A$ and $B$ are placed on opposite sides with respect to a black hole. During its trip, the light beam reaches the minimum radius $r_0$ and, once arrived at $B$, it gets reflected back to $A$.
The total time for the return trip of the light ray, denoted by $\Delta t$, can be obtained by integrating Eq.~\eqref{eq:time_radius_geod} and substituting Eq.~\eqref{eq:brel} we find
\begin{equation}
  \label{eq:time_elapsed}
  \Delta t=2\left(\int_{r_0}^{r_A}+\int_{r_0}^{r_B}\right) \frac{dr}{f(r)\sqrt{1-\dfrac{r_0^2}{r^2}\dfrac{f(r)}{f(r_0)}}} \, .
\end{equation}

The Shapiro time delay $\Delta t_{\rm d}$ may now be defined as the difference of $\Delta t$ and the duration of the return trip that the light would have taken if the gravitational source was not there to bend the spacetime. In other words, we have that the Shapiro time delay reads
\begin{equation}
  \Delta t_{\rm d}=\Delta t-2\left(\sqrt{r_A^2-r_0^2}+\sqrt{r_B^2-r_0^2}\right) \, .
\end{equation}
Note that in this definition we are using ``Schwarzschild-type coordinates''. For astrophysical applications different choices of coordinates might be preferred. 

Numerical results concerning the values of $\Delta t_{\rm d}$ for the geometry resulting from our metric function~\eqref{eq:quantumf} are provided in Fig.~\ref{fig:phi_time}, right panel.

\section{Quasinormal modes}
\label{sec-qnm}
It is now interesting to investigate some aspects of the spectrum of quasinormal modes for the quantum-corrected RN spacetime determined by the metric function~\eqref{eq:quantumf}. 

Since our spacetime is electrically charged, the electromagnetic and gravitational perturbations are coupled to each other and their analysis is quite cumbersome (see e.g.~\cite{Berti:2009kk}). So, for simplicity, we will just compute the quasinormal modes of scalar perturbations, that can be treated as the quasinormal modes of a minimally coupled scalar field $\Phi$ on our fixed background. In detail, a scalar field $\Phi$ of mass $m$ propagating on a background geometry determined by the metric function~\eqref{eq:quantumf} satisfies the Klein--Gordon equation 
\begin{equation}
\label{eq:KG}
\left(\Box-m^2\right)\Phi=0 \, , \qquad \Box := g^{\alpha \beta} \nabla_\alpha \nabla_\beta \, .
\end{equation}
Decomposing $\Phi$ into Fourier modes and spherical harmonics, i.e.
\begin{equation}
  \Phi(t,r,\theta,\phi)=\int d\omega\sum_{\ell,m}\,e^{-i\omega t}\,\frac{\psi(r)}{r}\,Y^{\ell}_m(\theta,\phi) \, ,
\end{equation}
Eq.~\eqref{eq:KG} can be recast (mode by mode) as a Schr\"odinger-like equation of the form 
\begin{equation}
\label{eq:schrodinger_scalar}
\frac{d^2\psi}{dr_*^2}+\left[\omega^2-V_0(r)\right]\psi=0
\end{equation}
with
\begin{equation}
\label{eq:scalarpot}
V_0(r)=f(r)\left(\frac{\ell (\ell+1)}{r^2}+\frac{f'(r)}{r}+m^2\right) ,
\end{equation}
where $r_*$ denotes the tortoise coordinate associated to the region outside the (outer) horizon. Considering the asymptotic behavior of the solutions of Eq.~\eqref{eq:schrodinger_scalar} near the horizon and at spatial infinity
\begin{equation}
\label{eq:asymptotics}
\psi(r_*)\sim
\begin{dcases}
Z_{\rm H}^{\rm (out)}\,e^{-i \omega r_*}+ Z_{\rm H}^{\rm (in)}\,e^{+i\omega r_*}\qquad&\text{for}\quad r\to R_{+}^{\rm q}\vspace{2mm}\\
Z_{\infty}^{\rm (out)}\,e^{+i\sqrt{\omega^2-m^2}\,r_*}+Z_{\infty}^{\rm (in)}\,e^{-i\sqrt{\omega^2-m^2}\, r_*}\qquad&\text{for}\quad r\to \infty
\end{dcases}
\end{equation}
then, the quasinormal mode spectrum is completely characterized by imposing the boundary conditions
\begin{equation}
\label{eq:inmodes}
Z_{\rm H}^{\rm (in)}=Z_{\infty}^{\rm (in)}=0 \, ,
\end{equation}
according to which no modes can come from inside the horizon or from infinity. With these boundary conditions it is found that the quasinormal mode frequencies $\omega$ are quantized in terms of an overtone number $n$ and the angular momentum number $\ell$.  

In the following we consider a scalar field with $m=0$ and set $R_{\rm M}=1$. We will calculate the quasinormal mode frequencies $\omega$ via a WKB method (see~\cite{Schutz:1985km}, and also~\cite{Berti:2009kk} for a comprehensive literature review), using Pad\'e approximants of order $[6/7]$. For more details on the code that was used in this work we refer the reader to \cite{WKBkonoplya2} by~\textsc{R. A. Konoplya et al.} (see also references therein for a link to the \textsc{Mathematica} package). 

We present results for the quasinormal modes only for the first values of $\ell$, and only for $n\leq\ell$, since this is the range where the WKB method is known to work best. In the tables, we denote by $(?)$ a result where the WKB approximation is not expected to be reliable, based on the error estimates returned by the ``automatic code'' discussed in~\cite{WKBkonoplya2}. This issue relates precisely to one of the intrinsic drawbacks of using the WKB method, i.e. the fact that higher-order corrections do not necessarily provide improved accuracy and normally lead to divergence
beyond a certain order (since the WKB expansion is asymptotic).

\bigbreak

\noindent Case 1: $R_{\rm Q}=0.1$ ($m=0$ and $R_{\rm M}=1$)
\vspace{0.2 cm}
\begin{center}
  \def\arraystretch{1.4}
  \begin{tabular}{c||c|c|c|}
    & Classical ($R_{\rm s}=0$) & Quantum ($R_{\rm s}=0.5$) & Quantum ($R_{\rm s}=0.7$) \\ \hline \hline
    $n=0,\ \ell=0$ & \hspace{0.5cm} $0.223 - 0.210 \,i\  (?)$ & \hspace{0.6cm}$0.222 - 0.207\, i\  (?)$ & \hspace{0.6cm}$0.220 - 0.183\, i\  (?)$\\ \hline
    $n=0,\ \ell=1$ & $0.590 - 0.196 \,i$& $0.590 - 0.194\, i$ & $0.592 - 0.183\, i$\\ \hline
    $n=1,\ \ell=1$ & $0.533 - 0.613     \,i$& \hspace{0.6cm}$0.533 - 0.611\, i\  (?)$ & \hspace{0.6cm}$0.532 - 0.561\, i\  (?)$\\ \hline
    $n=0,\ \ell=2$ & $0.974 - 0.194 \,i$& $0.974 - 0.193\, i$ & $0.977 - 0.182\, i$\\ \hline
    $n=1,\ \ell=2$ & $0.935 - 0.592 \,i$& $0.933 - 0.588\, i$ & \hspace{0.6cm}$0.938 - 0.551\, i\  (?)$\\ \hline
    $n=2,\ \ell=2$ & $0.868 - 1.019 \,i$& \hspace{0.6cm}$0.878 - 1.016\, i\  (?)$ & \hspace{0.6cm}$0.874 - 0.942\, i\  (?)$\\ \hline
  \end{tabular}
\end{center}
\vspace{0.6 cm}

\noindent Case 2: $R_{\rm Q}=0.3$ ($m=0$ and $R_{\rm M}=1$)
\vspace{0.2 cm}
\begin{center}
  \def\arraystretch{1.4}
  \begin{tabular}{c||c|c|c|}
    & Classical ($R_{\rm s}=0$) & Quantum ($R_{\rm s}=0.5$) & Quantum ($R_{\rm s}=0.7$) \\ \hline \hline
    $n=0,\ \ell=0$ & \hspace{0.5cm} $0.237 - 0.212 \,i\  (?)$ & \hspace{0.6cm}$0.235 - 0.201\, i\  (?)$ & \hspace{0.6cm}$0.227 - 0.162\, i\  (?)$\\ \hline
    $n=0,\ \ell=1$ & $0.627 - 0.198 \,i$& $0.631 - 0.189\, i$ & $0.648 - 0.163\, i$\\ \hline
    $n=1,\ \ell=1$ & $0.575 - 0.619     \,i$& \hspace{0.6cm}$0.578 - 0.500\, i\  (?)$ & \hspace{0.6cm}$0.530 - 0.565\, i\  (?)$\\ \hline
    $n=0,\ \ell=2$ & $1.035 - 0.197 \,i$& $1.041 - 0.188\, i$ & $1.068 - 0.161\, i$\\ \hline
    $n=1,\ \ell=2$ & $0.999 - 0.600 \,i$& $1.004 - 0.569\, i$ & \hspace{0.6cm}$1.041 - 0.491\, i\  (?)$\\ \hline
    $n=2,\ \ell=2$ & $0.938 - 1.028 \,i$& \hspace{0.6cm}$0.945 - 0.970\, i\  (?)$ & \hspace{0.6cm}$0.985 - 0.831\, i\  (?)$\\ \hline
  \end{tabular}
\end{center}
\vspace{0.6 cm}

In full analogy with the results of~\cite{antonelli2025quasinormalmodescoherentquantum}, the quasinormal mode frequencies seem to slightly deviate from those of the corresponding solution in General Relativity, i.e. the RN black hole, and the discrepancy increases, as expect, as we increase the size of the quantum core $R_{\rm s}$. Notably, we again find that the imaginary part of $\omega$ for the quantum-corrected geometry is smaller than that of the corresponding classical spacetime, thus leading to longer decay times for such modes, which increase with the size of the core. Conversely, the real part of $\omega$ remains mostly unaffected.

\section{Conclusions}
\label{sec-conclusions}
In this work we have improved upon a previous investigation of electrically charged coherent quantum black holes. Specifically, taking advantage of an alternative UV regularization procedure proposed in~\cite{Feng:2024nvv}, we have been able to remove the {\em spurious oscillations} found in the quantum-corrected RN geometry derived in~\cite{Casadio:2022ndh}. Interestingly, our novel quantum-corrected RN geometry given by
\begin{equation*}
    ds^2=-f(r)dt^2+\frac{dr^2}{f(r)}+r^2d\Omega_{(2)}^2, \qquad f(r) = 1-\frac{R_{\rm M}}{r}\erf{\hspace{-1mm}\left(\frac{r}{R_{\rm s}}\right)}+\frac{R_{\rm Q}^2}{r^2}\frac{2r}{R_{\rm s}}\F\hspace{-1mm}\left(\frac{r}{R_{\rm s}}\right) \, ,
\end{equation*}
(with $R_{\rm s}$ the size of the quantum core, $R_{\rm M}\equiv2G_{\rm N}M$, and $R_{\rm Q}\equiv\sqrt{G_{\rm N}}Q$) shares the same general features with the solution in~\cite{Casadio:2022ndh}, thus suggesting that these properties are truly regularization-independent, while loosing the problematic unphysical oscillations in the quantum potential.

The effective stress-energy tensor for this new geometry is described in terms of an anisotropic fluid whose components are locally integrable, with respect to the volume measure, for $r>0$. 

The causal structure of our proposed quantum geometry is also much richer than previously thought, allowing for both extremal and non-extremal scenarios as well as single horizons with multiplicity one. Notably, for the extremal case, the bound set by the weak gravity conjecture is satisfied for all the relevant values in the parameter space. Furthermore, from Fig.~\ref{fig:horizons_RN} it is easy to see that the novel geometry proposed here, similarly to the one presented in~\cite{Casadio:2022ndh}, allows for configurations free of Cauchy horizons. Such horizons are recovered for scenarios with quantum cores that are sufficiently small (i.e. closer to the classical limit of the approach).

Another interesting feature of the proposed geometry consists in the fact that if $R_{\rm s} = R_{\rm s}^\ast \equiv \sqrt{\pi} R^2_{\rm Q}/R_{\rm M}$, then the spacetime is free of singularities. If $R_{\rm s} \neq R_{\rm s}^\ast$, the spacetime features a much milder singular behavior, i.e. it contains an {\em integrable singularity} at $r=0$.

To complement our analysis in Sec.~\ref{sec-geo} we provide a discussion of some relevant observables stemming from the geodesics of our geometry, while in Sec.~\ref{sec-qnm} we discuss the spectrum of quasinormal modes of a scalar field perturbation. The results for the quasinormal modes, in particular, are consistent with those computed for the quantum-corrected Schwarzschild black hole~\cite{antonelli2025quasinormalmodescoherentquantum}. Specifically, we find that the quasinormal mode frequencies of our quantum-corrected RN black hole have imaginary parts that are slightly smaller than the corresponding counterparts from the RN black hole of General Relativity, while the real parts remain very close to the classical results.

The observables computed in this work depend explicitly on the UV cut-off given by the size of the quantum core $R_{\rm s}$. This is due to the fact that the geometry computed here is an effective geometry obtained by averaging over the would-be quantum-gravity region inside the black hole. Hence, the value of $R_{\rm s}$ should emerge from the underlying quantum gravity theory and cannot be computed within the coherent-state approach. To remain agnostic with respect to the more fundamental theory we can only set bounds on $R_{\rm s}$ so that the corresponding geometry is not plagued by standard instabilities occurring in most models of regular black holes, such as mass inflation. A theoretical indication that $R_{\rm s}$ should be a (macroscopically large)  fraction of the size of the event horizon, thus leading to an effective geometry devoid of Cauchy horizons and with an alleviated curvature singularity, was discussed in \cite{Casadio:2024lzd}.
\section*{Acknowledgments}
The authors would like to thank Roberto Casadio for valuable discussions.\\
The work of T.A. is supported by a doctoral studentship of the Science and Technology Facilities Council
(training grant No. ST/Y509620/1, project ref. 2917813).
The work of M.S. is supported by a doctoral studentship of the Science and Technology Facilities Council (training grant No. ST/X508822/1, project ref. 2753640).
A.G. is supported by the Italian Ministry of Universities and Research (MUR) through the grant ``BACHQ: Black Holes and The Quantum'' (grant no. J33C24003220006). The work of A.G. has also been carried out in the framework of activities of the National Group of Mathematical Physics (GNFM, INdAM).
\section*{Data Availability Statement}
This manuscript has no associated data.
\section*{Code Availability Statement}
This work takes advantage of codes that are already publicly available. Nonetheless, our code will be made available on reasonable request.
%
%
%
%
\bibliography{references}
\bibliographystyle{utphys}
\end{document}